\documentclass[aps,prl,twocolumn,superscriptaddress]{revtex4-1}
\usepackage{graphicx}
\usepackage{bm}
\usepackage{graphicx}

\begin{document}

\title{Competition between Coulomb and Symmetry Potential in Semi-peripheral Heavy-ion Collisions}

\author{Qianghua Wu} 
\affiliation{China Institute of Atomic Energy, Beijing 102413, P.R. China}
\affiliation{Department of Physics and Technology, Guangxi Normal University, Guilin 541004, China}

\author{Yingxun Zhang} 
\email{zhyx@ciae.ac.cn}
\affiliation{China Institute of Atomic Energy, Beijing 102413, P.R. China}

\author{Zhigang Xiao} 
\affiliation{Department of Physics, Tsinghua University, Beijing 100084, China}
\affiliation{Collaborative Innovation Center of Quantum Matter, Beijing 100084, China}

\author{Rensheng Wang} 
\affiliation{Department of Physics, Tsinghua University, Beijing 100084, China}

\author{Yan Zhang} 
\affiliation{Department of Physics, Tsinghua University, Beijing 100084, China}

\author{Zhuxia Li} 
\affiliation{China Institute of Atomic Energy, Beijing 102413, P.R. China}

\author{Ning Wang} 
\affiliation{Department of Physics and Technology, Guangxi Normal University, Guilin 541004, China}

\author{R. H. Showalter}
\affiliation{National Superconducting Cyclotron Laboratory Michigan State University, East Lansing, MI 48864, USA}
\affiliation{Department of Physics and Astronomy Michigan State University, East Lansing, MI 48864, USA}

\date{\today}




\begin{abstract}
The anisotropy of angular distributions of emitted nucleons and light charged particles for the asymmetric
reaction system, $^{40}$Ar+$^{197}$Au, at b=6fm and $E_{beam}$=35, 50 and 100MeV/u, are investigated by using
the Improved Quantum Molecular Dynamics model. The competition between the symmetry potential and Coulomb
potential shows large impacts on the nucleons and light charged particles emission in projectile and target
region. As a result of this competition, the angular distribution anisotropy of coalescence invariant Y(n)/Y(p)
ratio at forward regions shows sensitivity to the stiffness of symmetry energy as well as the value of Y(n)/Y(p).
This observable can be further checked against experimental data to understand the reaction mechanism and to
extract information about the symmetry energy at subsaturation densities.
\end{abstract}



\maketitle

\section{introduction}
The nuclear symmetry energy plays a crucial role for understanding not only nuclear structures and reactions but also astrophysical phenomena. However, theoretical predictions for the density dependence of symmetry energy of nuclear matter show large uncertainties away from the normal density \cite{Brown00,BALi08}. Many efforts have been devoted to probe and constrain the symmetry energy at both subsaturation and suprastaturation densities by analyzing observables from nuclear reactions, nuclear structures and neutron stars, such as: isospin diffusion\cite{Tsang01,HYWu02,Tsang04,LWChen05,TXLiu07,Sun10,Rizzo08}, double neutron to proton ratio (DR(n/p))\cite{Famiano06,BALi06,Zhang08,Tsang09,SKumar11,Zhang12,Zhang14,WJXie13}, light charged particle flow\cite{GCYong09,Kohley10,Giord10,Cozma11,YJWang14,Rizzo04}, $\pi^-/\pi^+$\cite{Xiao09,ZQFeng10,YGao13,JHong14,Xiao14}, neutron skin\cite{Warda10, LWChen10, Gaidarov12}, Giant Dipole Resonance and Pygmy Dipole Resonance \cite{Carbonne,Wieland,Piekar11} and masses of Isospin Analog State\cite{Danie09}, alpha decay\cite{JMDong12}, mass-radius relationship and gravitational waves from merging
neutron star binaries \cite{Stein12,Kentaro14}. Up to now, a consensus on the constraints of symmetry energy at subsaturation density \cite{Tsang12, Lattimer14} has been obtained. Nevertheless, differences among the constraints of symmetry energy obtained from different models or approaches still exist. Understanding these differences is a challenge in this field and stimulates nuclear physicists to propose new probes that are sensitive enough to further discriminate the stiffness of symmetry energy and to perform new experiments for a comprehensive understanding of the mechanism of neutron-rich heavy ion collisions. 

Some probes related to the slower reaction process for semi-peripheral collisions in asymmetric reaction systems near the Fermi energy have been investigated or measured to understand the reaction mechanism of neutron-rich heavy ion collisions(HIC) and to extract the information about symmetry energy\cite{Filippo,Hudan12, Amorini09, RSWang14,Rustto}. For example, in reference \cite{Filippo}, the isotopic composition of intermediate mass fragments (IMFs) emitted at mid-rapidity in semi-peripheral collisions of $^{124}$Sn+$^{64}$Ni was analyzed and it was found that the IMFs emitted in the early stage of the reaction show larger values of $<N/Z>$ and stronger angular anisotropy. The linear density dependence of symmetry energy has been obtained by comparing the data with SMF model. Calculations from transport model\cite{Zhang05} also show that the dynamical emission of nucleons and light charged particles (LCPs) for asymmetric reaction systems, such as $Y(n)/Y(p)$ ratios and $Y(t)/Y(^3He)$ ratios as a function of impact parameters, are also sensitive to the density dependence of symmetry energy due to the isospin migration/diffusion mechanism in neck region.
Very recently, R. S. Wang et al. analyzed the energy spectra of light charged particles in coincidence with fission fragments in $^{40}$Ar+$^{197}$Au at 35 MeV/u \cite{RSWang14}. The triggering condition on fission events in experiment has a bias on the semi-peripheral collisions. By defining the ratio of the isotopic yield (normalized to that of proton) at $80^\circ$ and $158^\circ$, $R_{iso}(X)=(\frac{\left[Y(X)/Y(p)\right]_{80}}{\left[Y(X)/Y(p)\right]_{158}})$, where X represents all species with Z=1 and 2, it is found that $R_{iso}$ increases with N/Z of the species, indicating a relatively high neutron richness of the emitted particles at smaller angle.
According to the studies in\cite{Colin03}, this result can be attributed to the hierarchical feature of the semi-peripheral collisions at Fermi energies, saying that the heavier fragments are faster than the lighter fragments along the beam direction and  the Coulomb potential is stronger near the projectile/target region compared to that at midrapidities. Thus, one can expect different Y(n)/Y(p) ratio at project/target region and midrapidities, naturally leading to a non-unity value of $R_{iso}$. The different form of symmetry energy may leads to different values of anisotropy of isospin contents, $R_{iso} (X)=R_X(\theta1)/R_X(\theta2)$, in the simulations. Furthermore, the non-unity value of $R_{iso}$ also suggests the reaction process is an nonequilibrium process. An investigation on angular distribution of the N/Z of the light particles may reveal the sequence of the particle emission with different isospin composition, which carries the information of the symmetry energy at subsaturation densities.

In this work, we mainly discuss the dynamical emission of nucleons and LCPs at early stages for asymmetric nuclear reaction system $^{40}$Ar+$^{197}$Au at semi-peripheral collisions by using the ImQMD05 code. In our simulations, the beam energy dependence of angular distribution of coalescence invariant neutron to proton ratios is also discussed from 35MeV/u to 100MeV/u. Based on our calculations, we propose that the anisotropy of angular distribution of coalescence invariant neutron to proton ratios may be used to understand the competition between the Coulomb and symmetry potential, and may be used for probing the isospin effects and form of symmetry energy.

\section{Brief introduction of improved quantum molecular dynamics model}
Within Improved quantum molecular dynamics model (ImQMD05 code) \cite{Zhang06, Zhang07,Zhang08}, nucleons are represented by Gaussian wavepackets and the mean fields acting on these wavepackets are derived from energy density functional with the potential energy $U$ that includes the Skyrme potential energy with just the spin-orbit term omitted:
\begin{equation}
         U=U_{\rho}+U_{md}+U_{coul}
\end{equation}
where $U_{\rho}$ is the Skyrme potential energy, $U_{md}$ is the momentum dependent potential energy, and $U_{coul}$ is the Coulomb energy. The nuclear contributions are represented in a local form with
\begin{equation}
        U_{\rho,md}=\int {u_{\rho,md}} d^3 r
\end{equation}
and,
\begin{widetext}
 \begin{equation}
    u_{\rho}= \frac{\alpha}{2} \frac{{\rho}^2}{\rho_0}
    + \frac{\beta}{ {\eta}+1} \frac{{\rho}^{{\eta}+1}}{\rho_0^{\eta}}
    + \frac{g_{sur}}{2\rho_0}{(\nabla{\rho})^2}
    + \frac{g_{sur,iso}}{\rho_0}  {[\nabla(\rho_n-\rho_p)]^2}
    + \frac{C_{s}}{2} {(\frac{\rho}{\rho_0})}^{\gamma_i} {\delta}^2 \rho
    + g_{\rho \tau} \frac{\rho^{8/3}}{\rho_0^{5/3}}
 \end{equation}
\end{widetext}
Here, $\delta$=$(\rho_n-\rho_p)$/$(\rho_n+\rho_p)$ is the isospin asymmetry, and $\rho_n$, $\rho_p$ are the neutron and proton densities, respectively. A symmetry potential energy density of the form $\frac{C_s}{2}(\rho/\rho_0)^{\gamma_i} \delta^2 \rho$ is used in the following calculations. The energy density associated with the mean-field momentum dependence is represented by
\begin{widetext}
 \begin{equation}
u_{md}=\frac{1}{2\rho_0} \sum_{N1}{\frac{1}{16{\pi}^6}} \int d^3 {p_1} d^3 {p_2} f_{N1}(\vec{p_1} )f_{N2}(\vec{p_2}) 1.57[ln(1+5\times 10^{-4} (\Delta{p})^2 ) ]^2
 \end{equation}
\end{widetext}
where $f_N$ are nucleon phase space distribution functions, $\Delta p$=$|\vec{p_1}-\vec{p_2}|$,
the energy is in MeV and momenta are in MeV/c. The resulting interaction between wavepackets is
described in Ref.\cite{Aiche87}. In this work, $\alpha$=-356 MeV, $\beta$=303 MeV, $\eta$=7/6,
$g_{sur}$=19.47 MeVfm$^2$, $g_{sur,iso}$=-11.35 MeVfm$^2$, $C_s$=35.19 MeV, and $g_{\rho,\tau}$=0 MeV.
The fragments are recognized by the isospin dependent minimum spanning tree method \cite{Zhang12R}.

\section{Results and discussion}
In this section, we examine the competition between the different form of symmetry potential and Coulomb potential
for peripheral collisions for $^{40}$Ar+$^{197}$Au by using the ImQMD model. The time evolution of density,
symmetry potential and single particle potential felt by neutron and
proton at overlapped region are analyzed. The competition can be clearly observed at projectile and target
region which leads to the different values of anisotropy of angular distribution of light particles for different form of symmetry potential.

\subsection{Reaction dynamics and time evolution of $V_{sym}^q$ and $V_q$ at neck region}
Figure 1 shows the density contour plots for $^{40}$Ar+$^{197}$Au reaction at the beam energy of 35MeV/u and impact parameter b=6fm. In the calculations, the projectile and target start to touch at around 50fm/c, and then the overlapped region is formed and it reaches maximum compression at around 90fm/c. After about 150fm/c, the overlapped region expands to lower density and ruptures into fragments. For semi-peripheral collisions we studied, the neck dynamics play important roles for the dynamical emission of LCPs\cite{Baran04}, and its yields and isospin contents for nucleons and LCPs are closely related to the symmetry potential in the participant region of the reaction. Thus, it is useful for us to understand how the symmetry potential and single particle potential for neutron and proton in the participant region evolve with time, because the single particle potential and symmetry potential will determine the yields and isospin contents of nucleons and LCPs emission.

\begin{figure}[htbp]
\centering
\includegraphics[angle=270,scale=0.30]{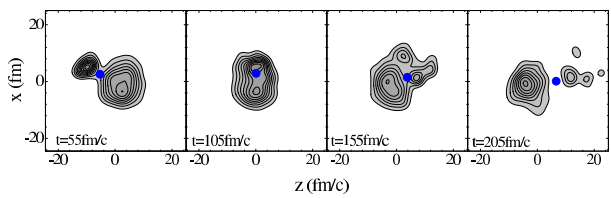}
\setlength{\abovecaptionskip}{30pt}
\caption{\label{fig:wide}(Color online) Time evolution of the density contour plots in the reaction plane
for $^{40}$Ar+$^{197}$Au collisions at $E_{beam}$=35 MeV/u and b=6fm from one typical event; the blue points represent the overlapped
region of projectile and target.}
\setlength{\belowcaptionskip}{0pt}
\end{figure}

Figure 2 (a) shows the time evolution of the symmetry potential $V_{sym}^q=\frac{C_s}{2}[(\gamma-1)u^\gamma\delta^2\pm 2u^\gamma\delta]$, $q=n,p$, at the overlapped region which is presented as blue points in Figure 1. The point is defined as the equal distance from the surface of the projectile and the target, and a spherical region with two times of nucleon radius, i.e. R=1.7fm, is used to approximately describe the center of overlapped region in the calculations. The solid lines are the results from $\gamma_i$=0.5 and dashed lines are for $\gamma_i$=2.0. The magnitude of the symmetry potential felt by neutron ($V_{sym}^n$) for $\gamma_i$=0.5 is obviously larger than that with $\gamma_i$=2.0 except for the time at around 80fm/c where $V_{sym}^n(\gamma_i=0.5)<V_{sym}^n(\gamma_i=2.0)$. The reason is that the compressed density is close and up to normal density at around 80fm/c, and its isospin asymmetry comes close to 0.2 due to isospin migration. Thus, the symmetry potential felt by neutron ($V_{sym}^n$) obtained with $\gamma_i$=2.0 becomes larger than that for $\gamma_i$=0.5 at $\rho>0.9\rho_0$ based on the formula of $V_{sym}^q$\cite{BALi97,Zhang05}. However, the magnitude of symmetry potential felt by proton for $\gamma_i$=2.0 is still smaller than that for $\gamma_i$=0.5 at $\rho\le1.1\rho_0$ for $\delta\sim0.2$, and it leads to the larger magnitude of symmetry potential felt by proton for $\gamma_i$=0.5 than that for $\gamma_i$=2.0 at different time for the beam energy of 35MeV/u. Since HIC observables obtained in the simulations are determined by the reaction process from early stages to late stages, one can expect that yield ratio of neutron to proton, i.e. Y(n)/Y(p), obtained with $\gamma_i$=0.5 is greater than that with $\gamma_i$=2.0 for neutron rich reaction system. Quantitatively, understanding the Y(n)/Y(p) ratios requires the knowledge of single particle potential felt by neutron and proton in the reaction system. Figure 2 (b) and (c) show the the single particle potential for neutron and proton for $\gamma_i$=0.5 and 2.0, respectively. The solid lines are for protons where the Coulomb contribution is included, i.e., $V_p+V_{Coul}$. The dashed lines are the nucleonic potential felt by the proton, i.e. $V_p$, without Coulomb potential. The dotted lines are the single particle potential for neutrons, i.e. $V_n$. Figure 2 (b) shows that $V_p+V_{Coul}$ is close to $V_n$ for $\gamma_i$=0.5. However, for $\gamma_i$=2.0, $V_p+V_{Coul}$ is obviously larger than $V_n$ after 120fm/c because symmetry potential becomes weak, especially at lower density. Thus, the Coulomb potential obviously moves the single particle potential felt by proton up to the strength by neutron. It may lead a higher yield for protons than neutrons, if the Coulomb repulsion is strong enough. Especially for asymmetric reaction systems at semi-peripheral collisions, one can expect to observe this competition between the Coulomb and symmetry potentials which influences the isospin contents for the emitted nucleons and LCPs, and hence, their angular distribution anisotropies show the sensitivity to the different forms of symmetry energy.
\begin{figure}[htbp]
\centering
\includegraphics[angle=270,scale=0.35]{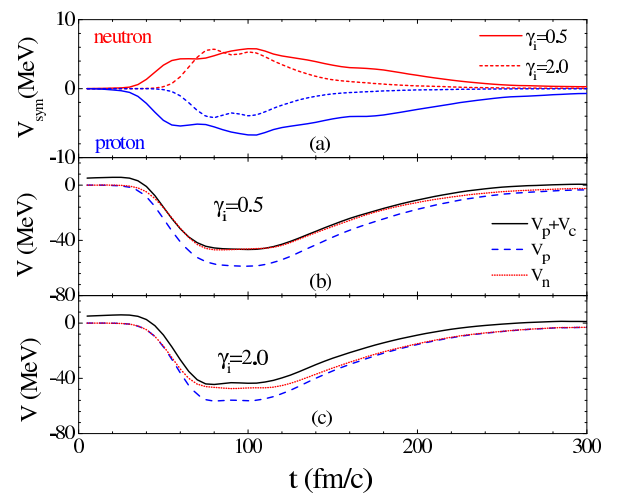}
\setlength{\abovecaptionskip}{40pt}
\caption{\label{fig2:vpot} (Color online) Time evolution of symmetry potential for neutron and proton obtained with $\gamma_i$=0.5 (solid lines) and 2.0 (dashed lines)(a); single particle potential for neutron and proton for $\gamma_i=0.5$ (b); and $\gamma_i=2.0$ (c) at the overlapped region for $^{40}$Ar+$^{197}$Au at b=6fm and $E_{beam}$=35MeV/u.}
\setlength{\belowcaptionskip}{20pt}
\end{figure}

\subsection{Angular distribution of the coalescence invariant yields for neutron and proton}
As is known, the absolute yields of LCPs are not very well reproduced due to the considerable deviation
of binding energy of light clusters between the QMD model and experimental data \cite{Neubauer99}.
Loosely bound clusters like $d$, $t$ are over predicted, whereas strongly bound clusters like $^4$He
is underpredicted due to its lower binding energy in the QMD model. Therefore, it is hard to draw a
firm conclusion by comparing the absolute number of these light fragments to data to extract the information
about physical quantities, such as the equation of state. One method to eliminate the problem related to
the absolute yield of LCPs is to introduce the coalescence invariant neutron and proton yields, and to
construct the related observables, such as Coalescence Invariant Y(n)/Y(p) (CI n/p ratio) ratio spectra\cite{Famiano06,Zhang08,Coupland14}.
In this paper, we investigate the angular distributions of CI n/p ratio by varying the stiffness of symmetry potential.
The angular distribution of coalescence invariant (CI) neutron and proton yields are constructed by adding
the neutrons and protons in light particles to free neutrons and protons at given $\theta_{c.m.}$ as follows:
\begin{eqnarray}
  \frac{dM_{n,CI}}{d\theta_{c.m.}}=\sum_{N,Z} N\cdot{\frac{dY(N,Z)}{d\theta_{c.m.}}}\\
  \frac{dM_{p,CI}}{d\theta_{c.m.}}=\sum_{N,Z} Z\cdot{\frac{dY(N,Z)}{d\theta_{c.m.}}},
\end{eqnarray}
the summation is over n, p, d, t, $^3$He, $^4$He and $^6$He in this work, and Y(N,Z) is the yield of fragments with charge number Z and neutron number N.

In order to understand the origin of angular distribution of CI yields for neutron and proton in asymmetric
reaction systems $^{40}$Ar+$^{197}$Au at 35MeV/u, we show the angular distribution of the yields for CI neutrons
and CI protons emitted from the projectile (line with circles) and the target (line with squares) at 150fm/c, 200fm/c and 350fm/c in Figure 3.
Left panels are the results for $\gamma_i$=0.5 (Fig.3(a) and (c)), and right panels are for 2.0
(Fig.3(b) and (d)).
The line with open (solid) symbols refers to the yields of CI protons in Fig.3 (a) and (b) (CI neutrons are in Fig.3(c) and (d)).
The calculated results
show that the yields of CI nucleons from target have a wide angular distribution and exhibit weak
anisotropy in the angular distributions. The peak of angular distribution appears
below $\theta_{c.m.}\sim 90^\circ$ when the time is earlier than $\sim$300fm/c, i.e., the nucleons
and LCPs prefer to emit at forward regions for asymmetric reaction systems because the neck dynamics
dominate the nucleons' and LCPs' emission at earlier stages. After then, the peak moves to the backward
regions at late stages because the evaporation from target-like fragments becomes more and more important.
The yields of CI nucleons from the projectile obviously peak around $\theta_{c.m.}\sim 25^\circ$ and
the corresponding angular position does not obviously vary with time. As shown in Figure 3, the nucleons
emitted at $\theta_{c.m.}<25^\circ$ are mainly from the projectile while those at $\theta_{c.m.}>60^\circ$ are
mainly from the target. The emission of nucleons within the angular gate $20^\circ<\theta_{c.m.}<60^\circ$
corresponds to the compressed overlap region between the projectile and target. The anisotropy of angular
distribution of CI nucleons is due to the asymmetry of projectile and target in semi-peripheral collisions.
Furthermore, one also observes that the yields of neutrons are obviously larger than that for protons
after 150fm/c for $\gamma_i$=0.5. But for $\gamma_i$=2.0, the yields of protons from target are always
higher than that for neutrons when $\theta_{c.m.}>60^\circ$. The reason is that the strength of
the symmetry potential is weak for $\gamma_i$=2.0 at low density regions where many nucleons are
emitted, and the target-like fragments provide stronger Coulomb repulsive. Thus, the yields of
protons become larger than neutrons for $\gamma_i$=2.0. For the nucleons from the projectile, one
also observe that yields of protons are greater than neutrons before 300fm/c, but it finally turns
over. After 300fm/c, the yields of protons and neutrons become similar because of the weaker
Coulomb potential in the projectile region.

\begin{figure}[htbp]
\begin{center}
\includegraphics[angle=270,scale=0.35]{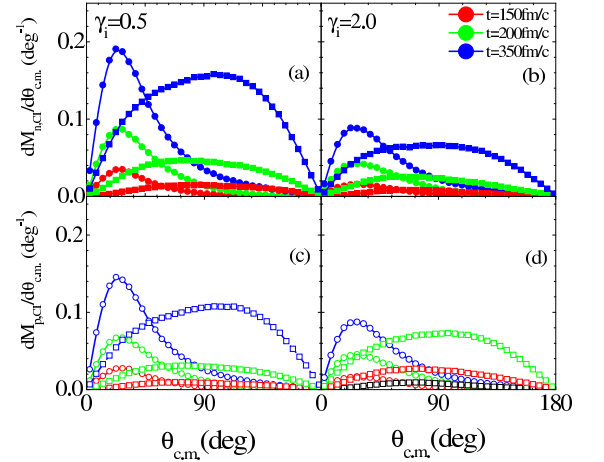}
\setlength{\abovecaptionskip}{50pt}
\caption{\label{fig:wide} (Color online) The angular distribution of the yields for CI neutrons ((a) and (b)),
and CI protons ((c) and (d)) emitted from projectile (line with circle symbols, left curves in each panel) or target
(line with square symbols, right curves in each panel)
at 150fm/c, 200fm/c and 350fm/c. Left panels are for $\gamma_i=0.5$ and right panels are for 2.0. }
\end{center}
\setlength{\belowcaptionskip}{0pt}
\end{figure}

Figure 4(a) shows the angular distributions of yields of CI neutrons (solid symbols) and protons (open symbols)
for $\gamma_i$=0.5 (red squares) and $\gamma_i$=2.0 (blue circles) at the stop time of simulations, t=400fm/c,
for $E_{beam}$=35MeV/u and b=6fm. In general, the yields of nucleons obtained with $\gamma_i$=0.5 are greater
than that for $\gamma_i$=2.0 due to its stronger symmetry energy at subsaturation densities. For the isospin
contents, the yields of neutrons are always greater than the yields of protons at whole angular range
for $\gamma_i$=0.5. Similar to the results at 35MeV/u, the yields of neutrons are greater than protons
for the beam energy at 50 (Figure 4 (b)) and 100MeV/u (Figure 4 (c)). But for the $\gamma_i$=2.0, a different
behavior appears in the yields of neutrons and protons. At the beam energy of 35MeV/u, the yields of neutrons
become less a little bit than the yields of protons at $\theta_{c.m.}>60^\circ$ for $\gamma_i$=2.0. It is the
result of the competition between the Coulomb potential and symmetry potential for asymmetric reaction system
at semi-peripheral collisions as mentioned before. At the beam energy of 100MeV/u,
the reaction is more violent and a larger part of the colliding system is dissociated into gases
(nucleons and light particles), and thus more neutrons originally bounded in the system are released
and outnumber the protons for the whole angular region, as shown in Figure 4(c).
\begin{figure}[htbp]
\centering
\includegraphics[angle=270,scale=0.35]{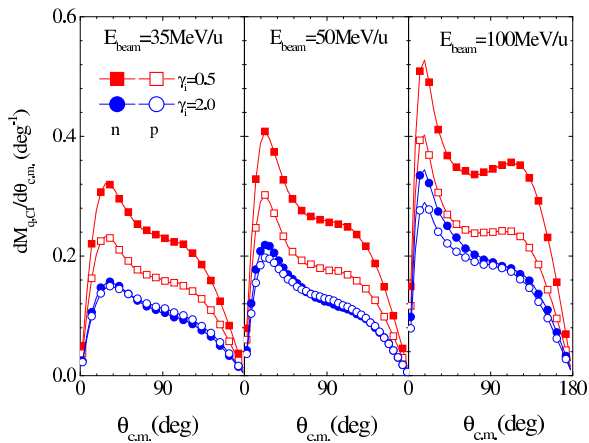}
\setlength{\abovecaptionskip}{40pt}
\caption{\label{fig:wide} (Color online) Angular distribution of the yields for CI neutrons (solid symbols) and
CI protons (open symbols) emitted at t=400fm/c obtained with $\gamma_i=0.5$ (red symbols) and 2.0 (blue symbols).
The left panel is the result at the beam energy of 35MeV/u, middle panel is for 50MeV/u and right panel is the
result for 100MeV/u. The impact parameter is 6fm.}
\setlength{\belowcaptionskip}{20pt}
\end{figure}

\subsection{Correlation between the symmetry energy and angular distribution of CI n/p ratios}
Consequently, we construct the angular distribution of CI n/p ratios, i.e. $\frac{dM_{n,CI}}{d\theta_{c.m.}}/\frac{dM_{p,CI}}{d\theta_{c.m.}}$
 ratios, to explore
the information of reaction mechanism and symmetry energy. As shown in Figure 5,
the CI n/p ratios obtained with $\gamma_i$=0.5 are greater than that with $\gamma_i$=2.0 for the beam
energy we studied, because the symmetry energy for $\gamma_i$=0.5 is stronger than $\gamma_i$=2.0 at subsaturation
density. The sensitivity of CI n/p ratios to $\gamma_i$ becomes large at the target region where the  charge number
of fragment is large, and it is consistent with the results in \cite{BALi05} where the influence of symmetry potential
and in-medium NN cross section on the free nucleon n/p ratio were studied for the central collision
of $^{100}$Zn+$^{40}$Ca at 200MeV/u.

In addition to the values of CI n/p ratio, analyzing the $\theta_{c.m.}$ dependence on the CI n/p ratio will provide more information for revealing the reaction mechanism on the competition between the Coulomb and symmetry potential. For $E_{beam}$=35MeV/u, the CI n/p ratios show a different $\theta_{c.m.}$ dependence for the different forms of symmetry potential. In case of $\gamma_i$=0.5, the CI n/p ratios slightly increase as a function of $\theta_{c.m.}$. It is the result of the isospin asymmetry changing from $\delta_{proj}$=0.10 at the projectile region to $\delta_{tar}$=0.19 at the target region since the single particle potentials of neutrons and protons are close, as shown in Figure 2 (b). However, in case of $\gamma_i$=2.0, the CI n/p ratios obviously decrease with $\theta_{c.m.}$, and the CI n/p ratios at backward are smaller than that at forward regions.
Furthermore, we also analyze the angular distributions of the CI n/p ratios at beam energy for 50 and 100MeV/u. The CI n/p ratios increase as the beam energy increases except for the CI n/p at forward region and $\gamma_i$=0.5.
The increasing behavior of the CI n/p ratios as the beam energy increases are from the enhancement of the symmetry potential at larger compressed density which can be achieved with higher beam energies. However, CI n/p ratios at forward regions for $\gamma_i$=0.5 show opposite behaviors, i.e. the CI n/p ratios decrease with the beam energy increasing. It is the effects of the finite size of the reaction systems. The final values of the CI n/p ratios from heavy ion collisions not only depend on the symmetry potential, but also depend on the yield of nucleons for finite reaction system. Since there are too many nucleons and LCPs emitted from the projectile for $\gamma_i$=0.5 at $E_{beam}$=100MeV/u, the CI n/p ratios at forward regions tend to decrease to the initial N/Z value of projectile, $N/Z_{proj}$=1.22, and is obviously smaller than that obtained at 35MeV/u.

To check the smearing effects of impact parameters on the sensitivity of this observable to $\gamma_i$,
calculations for b=4 and 8fm have also been performed with ImQMD. The calculations show that the CI n/p ratio
insignificantly depends on the impact parameters from b=4 to 8fm at the beam energy of 35, 50 and 100 MeV/u,
because the value of CI n/p ratio reflects the symmetry energy information of reaction system which is not varied
dramatically from b=4 to 8 fm for $^{40}$Ar+$^{197}$Au. For its insensitivity to the changing of the impact parameter,
the CI n/p ratio for mid-peripheral collisions shall be a valid probe to the symmetry energy.
\begin{figure}[htbp]
\centering
\includegraphics[angle=270,scale=0.26]{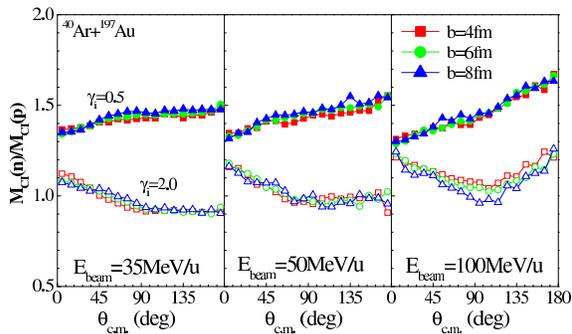}
\setlength{\abovecaptionskip}{50pt}
\caption{\label{fig:wide} (Color online) Left panel shows the calculated results for CI n/p ratios as a function of $\theta_{c.m.}$ for b=4, 6 and 8fm at $E_{beam}$=35MeV/u. The solid symbols are for $\gamma_i=0.5$ and open symbols are for $\gamma_i=2.0$. Middle panel is the results for 50MeV/u and right panel is for 100MeV/u.}
\setlength{\belowcaptionskip}{50pt}
\end{figure}

Focusing on the anisotropy of angular distributions of the CI n/p ratios and its sensitivity
to $\gamma_i$ (or the slope of symmetry energy, $L=3\rho_0\frac{\partial S(\rho)}{\partial \rho}|_{\rho_0}$,),
we analyzed $R_{iso1}=R_{CIn/p} (20^\circ)/R_{CIn/p} (90^\circ)$ and $R_{iso2}=R_{CIn/p} (90^\circ)/R_{CIn/p} (160^\circ)$,
which makes the experimental discrimination easier, for b=6fm at different beam energies.
The calculations show $R_{iso1}$ increases by $\sim$20\% when $\gamma_i$ varying
from $0.5$ to $1.5$ for $E_{beam}$=35MeV/u. For $R_{iso2}$, which is constructed from the backward regions,
its values weakly depend on the symmetry energy for the beam energies we studied. Similar behaviors have also
been found for 50 and 100MeV/u. These calculations illustrate that $R_{iso1}$ is strongly correlated to the
stiffness of symmetry energy, and it would be helpful for us to extract the information of symmetry energy
and to understand the reaction mechanism as well as the spectra of CI n/p ratios if more complete data are measured.

\begin{figure}[htbp]
\centering
\includegraphics[angle=270,scale=0.35]{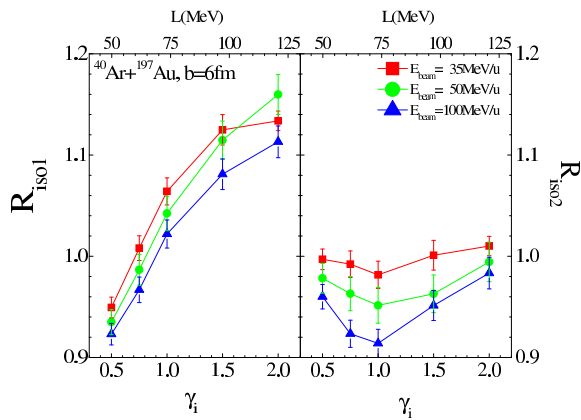}
\setlength{\abovecaptionskip}{50pt}
\caption{\label{fig:wide} (Color online) The calculated results of $R_{iso1}$ (left panel) and $R_{iso2}$ (right panel) as a function of $\gamma_i$ for $^{40}$Ar+$^{197}$Au at b=6fm and the beam energy is 35, 50 and 100 MeV/u.}
\setlength{\belowcaptionskip}{50pt}
\end{figure}

\section{summary}
In summary, the semi-peripheral heavy-ion collisions of $^{40}$Ar+$^{197}$Au at $E_{beam}$ =35, 50 and 100MeV/u were studied by means of the improved quantum molecular dynamics model (ImQMD05).
The competition between the symmetry potential and Coulomb potential is discussed for the asymmetric reaction system. The calculations show that the impacts of the competition between Coulomb and symmetry potential on the heavy ion collision observables, CI n/p ratio, can be observed at the projectile and target regions. In case of $\gamma_i$=2.0, the yields of protons are larger than neutrons in the target region for beam energy at 35MeV/u due to its stronger Coulomb potential and weaker symmetry potential. As the beam energy increasing up to 100MeV/u, the reaction is more violent and a larger part of the colliding system is dissociated into gases (nucleons and light particles), and thus more neutrons originally bounded in the system are released and outnumber the protons. For the $\gamma_i$=0.5, the yields of neutrons are obviously greater than protons due to its stronger symmetry potential at subsaturation density.
The competition between the Coulomb and symmetry potential causes the anisotropy of angular distribution of coalescence invariant n/p ratio, $R_{iso1}$, which is accordingly sensitive to the stiffness of symmetry energy as well as the values of CI n/p ratio. For $\gamma_i$=2.0, the CI n/p ratios decrease with angle increasing,
and $R_{iso1}=1.14\pm0.01$. For $\gamma_i$=0.5, the CI n/p ratios slightly increase with angle because the single particle potential felt by proton is close to neutron, and $R_{iso1}=0.94\pm0.01$. The values of $R_{iso1}$ increase by about $\sim$20\% when
$\gamma_i$ varies from 0.5 ($L$=51MeV) to 1.5($L$=104MeV). The sensitivity of $R_{iso1}$ to $\gamma_i$ remains at all the beam energies we studied. It is of high interest to measure the isospin composition of the light particles over a wide angular range in further experiments at Fermi energies to gain comprehensive understanding on the reaction mechanism as well as on the the symmetry energy behavior at subsaturation densities, and the symmetry energy in warm dilute matter.



\begin{acknowledgements}
This work has been supported by the 973 Program of China No. 2013CB834404,
the Chinese National Science Foundation under Grants (11475262, 11275052, 11422548, 11375062,11375094,11365004)
and CUSTIPEN (China\text{-}US Theory Institute for Physics with Exotic Nuclei) under Department of
Energy Grant No. DE-FG02-13ER42025. One of the authors Yingxun Zhang thanks Prof. J.R. Stone
helpful discussions on the symmetry energy in nuclei.
\end{acknowledgements}

\end{document}